\documentclass[review]{elsarticle}
\usepackage[legalpaper, margin=1in]{geometry}
\usepackage{booktabs}
\usepackage{lineno,hyperref}
\usepackage{amsmath}
\modulolinenumbers[5]
\usepackage{float}
\usepackage{graphicx}
\usepackage[caption=false]{subfig}
\journal{International Communications in Heat and Mass Transfer}

\begin{document}

\begin{frontmatter}

\title{Machine learning assisted modeling of thermohydraulic correlations for heat exchangers with twisted tape inserts}

\author{J.P. Panda} \author{B. Kumar \fnref{myfootnote}}
\author{A.K. Patil} \author{M. Kumar}
\address{Department of Mechanical Engineering,
DIT University, Dehradun, Uttarakhand, India}
\fntext[myfootnote]{corresponding author: bipin.kumar@dituniversity.edu.in}

\begin{abstract}
This article presents the application of machine learning (ML) algorithms in modeling of the heat transfer correlations (e.g. Nusselt number and friction factor) for a heat exchanger with twisted tape inserts. The experimental data for the heat exchanger at different Reynolds numbers and twist ratios were used for the correlation modeling. Three machine learning algorithms: Polynomial Regression (PR), Random Forest (RF), and Artificial Neural Network (ANN) were used in the data-driven surrogate modeling. The hyperparameters of the ML models are carefully optimized to ensure generalizability. The performance parameters (e. g. $R^2$ and $MSE$) of different ML algorithms are analyzed. It was observed that the ANN predictions of heat transfer coefficients outperform the predictions of PR and RF across different test datasets. Based on our analysis we make recommendations for future data-driven modeling efforts of heat transfer correlations and similar studies.            
\end{abstract}

\begin{keyword}
Machine learning\sep Heat transfer correlation\sep Polynomial regression \sep Random forest \sep Artificial neural network
\end{keyword}

\end{frontmatter}


\section{Introduction}
The heat transfer in different heat exchangers is largely affected by their geometries and flow conditions. The heat transfer phenomenon in those heat exchangers is generally studied by calculating the heat transfer coefficient, Nusselt number, and friction factor. Traditional data-driven approaches like linear regression may fit simple datasets with a limited number of input parameters. However, when there are a large number of input parameters and the relationship between the features and labels is complicated, more complex algorithms are required to model the relationships between the input features and the target outputs. Examples of such complications include a heat exchanger with a number of twisted tapes of varying twist ratio, different Reynolds numbers and number of holes on the twisted tape and having a porosity of varying magnitude, etc. Such irregularities in the flow may develop certain non-linearity in the flow field and result in complex heat transfer. The heat transfer coefficients for the above-mentioned complex heat transfer problems may not be accurately modeled with traditional data-driven models and require advanced ML algorithms.

Currently, there is considerable interest regarding the application of ML techniques in modeling different processes of scientific interest starting from robotics, manufacturing to healthcare and control engineering. In the field of fluid mechanics, researchers are applying ML and deep learning algorithms to develop new turbulence models and even predicting complex flow fields accurately. Such accurate prediction was not possible using traditional models that relied on physics-based precepts. In manufacturing, researchers are applying ML algorithms for modeling different machining responses in terms of various machining parameters and utilizing the ML-based surrogate to optimize the machining process parameters. 

Various researchers have applied ML algorithms in modeling and predicting different problems of thermal engineering and heat transfer (\citet{hughes2021status}). \citet{baghban2019sensitivity} used artificial neural network (ANN), support vector machine (SVM), and adaptive neuro-fuzzy inference system (ANFIS) for modeling the Nusselt number (Nu) in terms of Prandtl number, volumetric concentration, and helical number as input parameters of the ML models for a helically coiled tube heat exchanger. The SVM regressor was found to be the best ML model in predicting the Nusselt number. \citet{warey2020data} utilized ANN, random forest (RF), and linear regression with stochastic gradient descent to model and predict the equivalent homogeneous temperature for all passengers and the cabin temperature averaged by its volume. The data required for developing the ML models have obtained from the Computational Fluid Dynamics (CFD) simulations of flow in the vehicle cabin model. The errors of predictions of the ML models were found to be less than 5\% of the actual values. \citet{wei2020machine} studied thermal transport phenomenon in porous media using ML algorithms.  
The effective thermal conductivity was predicted in terms of different structural features. \citet{zhu2021machine} utilized ANN models to predict heat transfer coefficient for two-phase boiling and heat transfer in mini channels. The data required for model development were obtained from experiments on boiling and condensation of a refrigerant. The mean absolute error of prediction was found to be within 10\%. \citet{kim2021prediction} predicted the heat flux for narrow rectangular channels for steady-state conditions using different ML algorithms. The data required for training an ANN was obtained from critical heat flux correlations. The average error of prediction of the ANN was found to be $3.65$\%. \citet{wang2019performance} used AdaBoost and SVM regressor for prediction of COP and heating capacity of a heat pump. The different features are outdoor and indoor temperature, indoor mass flow rate, compressor speed, and two factors for controlling devices. It was observed that the ML predictions error is 8.5\% of the actual values. \citet{akay2021modeling} modeled the total heat transfer coefficient different ML algorithms such as multilayer perception (MLP), support vector machine (SVM), and M5P model tree and also proposed an equation based on the NTU method. The SVM predictions were found to be better in comparison to the other methods.
\citet{asgari2021gray} developed a gray box model by combining an ANN model with thermo-fluid transport equations to predict transient temperatures in server CPUs. The ANN mainly predicts the pressure, which can be used as input to the thermo-fluid transport equations from which the temperature field can be predicted. The gray-box model outperforms all the existing physics-based and black box models in predicting the temperature field. \citet{kwon2020machine} developed multi-variable heat transfer correlations for heat transfer in a cooling channel with ribs of variable roughness. An RF algorithm was trained for predicting the heat transfer correlation in terms of different rib geometries. \citet{zhu2021heat} used an ANN with two hidden layers to predict supercritical carbon dioxide heat transfer in a vertical tube. The ANN model was able to predict supercritical wall temperature with an acceptable level of accuracy. \citet{alizadeh2021machine} applied ANN on numerically predicted data of flow and heat transfer of nanofluid $(Al_{2}O_{3}-Cu-water)$ flowing along a cylinder embedded in porous media to model the Thermo-hydraulic field. They also applied particle swarm optimization to develop correlations for shear stress and Nusselt number.           

In this article, we investigate the efficacy and efficiency of different ML algorithms in modeling correlations for Nusselt number and friction factor in terms of flow and geometrical features for heat exchangers with twisted tape inserts. The flow and geometric features are Reynolds number and twist ratio of the twisted tape respectively. The data required for the development of ML models were collected from the experimental dataset of Nusselt number and friction factor from the experiments of \citet{kumar2019effect}. A brief discussion on experimental methods for data collection is presented in section \ref{sec:exp}. Two different train-test datasets are prepared by considering the variability of Reynolds number and twist ratio. In the training phase of the ML algorithms, the hyperparameters of the models are optimized. Once the ML models are developed using the experimental dataset, those are applied over test data to check their predictive capability. The predictive capability of ML-based surrogate models was assessed in terms of $R^2$ and $MSE$ of testing. Based on the predictive capability the best ML model for modeling the heat transfer correlations is recommended.

The rest of the article is arranged as follows: in section 2 the methodology adopted in the ML model development is presented. In section 3 the experimental data collection methodology is discussed. In section 4 the different ML algorithms are presented. In sections 5 and 6, the training and testing of the ML models are discussed and finally, in section 7, the conclusions are discussed.

\section{Methodology adopted}
\begin{figure}
\begin{centering}
\includegraphics[width=0.35\textwidth]{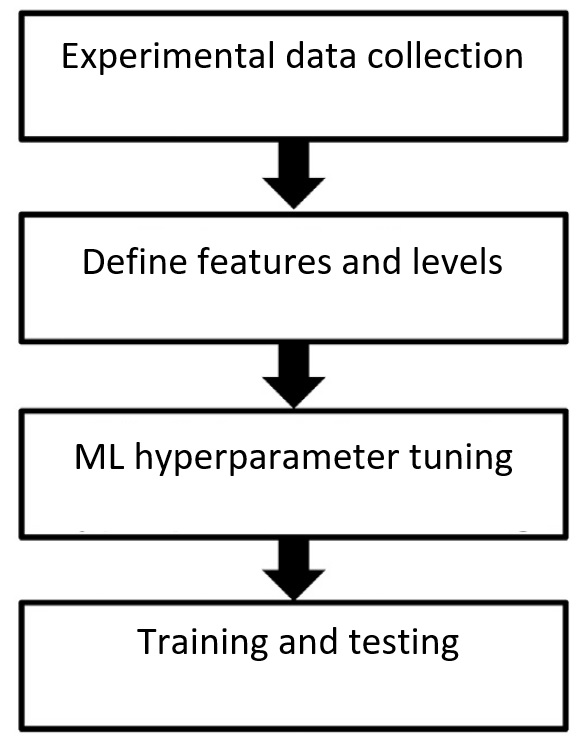}
\caption{The flow chart of ML model development.\label{fig:ML_flow}}
\end{centering}
\end{figure}

In figure \ref{fig:ML_flow} the detailed methodology of modeling the heat transfer correlations in terms of the flow and geometrical parameters (input parameters) are presented. The first stage of any ML-based model development is the collection of data and preparation of a suitable dataset by varying different input parameters. Any data-driven model is bounded by the data that it is trained on. The data for heat transfer correlations can either be prepared by conducting experiments or by performing numerical simulations. In this work, the experimental data collected at the Department of Mechanical Engineering, DIT University is used for ML model development. The details of experimental methodology and techniques of velocity and temperature measurement will be presented in the subsequent sections. Once the experimental data are collected, it is required to process those for use in the development of ML models. The next step is defining the features and levels. In ML-based modeling, the features and levels are set of input and output parameters respectively. Features are characteristic parameters defining and describing each sample. Once the features and levels are decided, it is required to choose a suitable ML algorithm for developing the surrogate model. The next step is to divide the dataset into training, validation, and testing sets. Then the dataset can be imported into the software, for the development of the ML model. The very first step of ML model training is tuning the hyperparameters. The predictive capability of ML models largely depends upon the fine-tuning of hyperparameters. There are different techniques by which the hyperparameters of ML algorithms can be optimized. Those are manual search, grid search, random search, and Bayesian optimization. In this work, we have optimized the hyperparameters using the manual approach. Once the hyper-parameters are fixed, the model can be applied for testing and prediction.

\section{Experimental data for heat exchanger with twisted tape inserts}
\label{sec:exp}
\begin{figure*}
\begin{centering}
\includegraphics[width=0.8\textwidth]{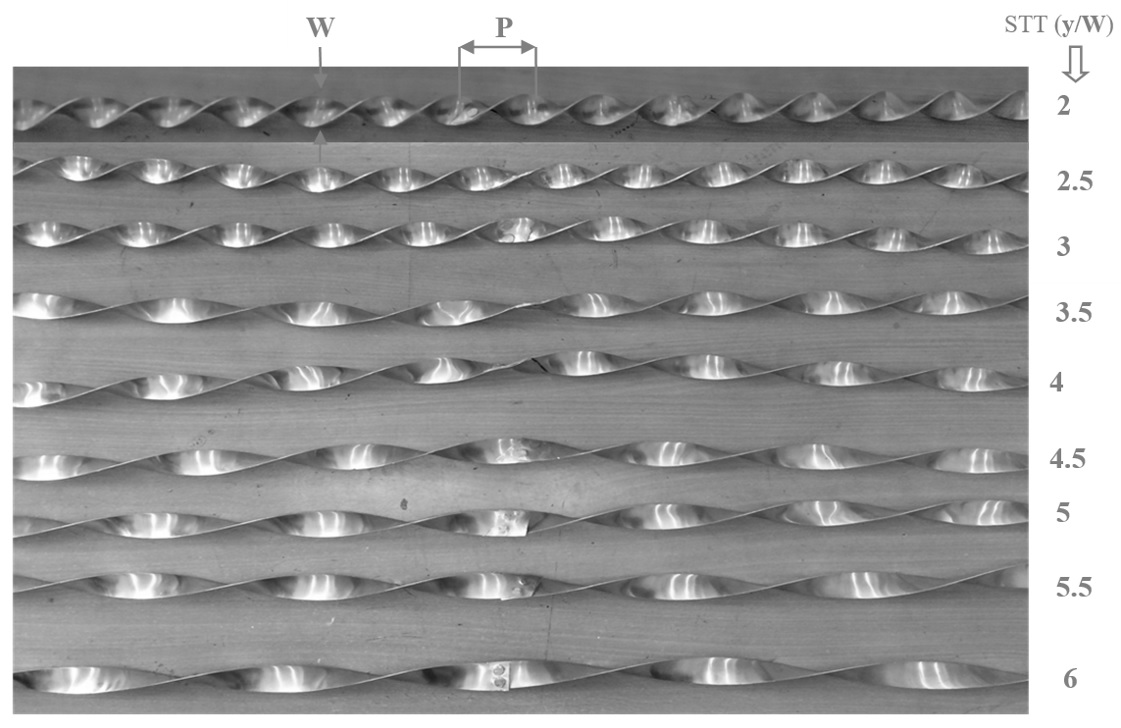}
\caption{Different twisted tapes used in experimental data collection \cite{kumar2019effect} for ML based surrogate development.\label{fig:ML_twisted}}
\end{centering}
\end{figure*}

The experimental data of \citet{kumar2019effect} for Nusselt number and friction factor at different Reynolds numbers and twist ratio is used for ML-based surrogate model development. The different twisted tapes used in experimental data generation are shown in figure \ref{fig:ML_twisted}. With the decrease in twist ratio, the turbulence level in the flow field increases. The increased turbulence level in the flow field signifies better mixing of fluid layers and enhanced heat transfer. The experimental data were collected for a cylindrical heat exchanger with a length of 1 meter and a diameter of 0.032 meters. A constant heat flux of 1000 $W/m^2$ was applied over the wall and the flow velocities at the inlet were varied to obtain different Reynolds numbers. The Nusselt number was calculated from the estimated values convective heat transfer rate, which was measured by recording the tube surface, inlet, and outlet temperatures, and the friction factor was calculated from the measured pressure drop between the inlet and outlet of the tube.

The Reynolds number, Nusselt number and friction factor can be defined as follows:

\begin{equation}
Re=\frac{\rho v D}{\mu}
\end{equation}

\begin{equation}
Nu=\frac{h D}{k}
\end{equation}

\begin{equation}
f=\frac{\Delta p}{(L/D)(\rho v^2/2)}
\end{equation}
In the above mentioned equations, $\rho$ is density, $D$ is the hydraulic diameter $k$ is the thermal conductivity, $h$ is the heat transfer coefficient, $L$ is the length of the tube and $v$ is the velocity of flow in the tube.  

\subsection{Description of the train-test and prediction datasets}
For the development of the ML-based surrogate models for prediction of $Nu$ and $F$, we have prepared two datasets $(D_1 and D_2)$ with different combinations of $Re$ and $t$. The prediction data for dataset $D_1$ has all the $Re$ variations and constant $t=3$ and The prediction data for dataset $D_2$ has all the $t$ variations and constant $Re=10449$. All the other data except prediction data were used for the training and testing of the ML models. The testing data has 20\% data from the train-test dataset, those were considered arbitrarily. The prediction data were not used in any phase of the model development, rather those were kept separate from the train-test data for testing the predictive capability of the developed ML models.     

\section{ML algorithms for thermohydraullic correlation modelling}

\subsection{Artificial Neural Networks (ANN)}
Artificial Neural Networks (ANN) are a popular machine learning algorithm and are also known as multi-layer perception. Those are inspired by biological neural networks. The most basic unit of ANN is a perceptron, also referred to as a neuron. A neuron takes input and performs a sequence of 2 operations on the input: a linear transform represented by a matrix multiplication and a non-linear transform represented via an activation. In a neural network, these neurons are arranged in groups referred to as layers of the network. For complex models, the neural network may consist of several such layers. The output of the layer $l$ of the network depends on the input from the layer $l-1$ via:      

\begin{equation}
\begin{aligned}
& q_{i}^{l}=\eta(\sum_j W_{ij}^{l}q_{j}^{l-1}).
\end{aligned}
\end{equation}
where $\eta$ is the activation function and the $W_{ij}$ are the weights associated with the layer. Training the neural network involves using the data to infer the best values of these weights. The gradients of the loss with respect to the weights of the ANN are calculated using back propagation algorithm (\citet{pandamodeling}). The gradient descent method can be used to to minimize the loss. Some of the activation functions, which can be used in MLP are sigmoid $\eta(\beta)=1/(1+e^{-\beta})$, hyperbolic tangent(tanh) $\eta(\beta)=(e^{-\beta}-e^{-\beta})/(e^{-\beta}+e^{-\beta})$ and RELU $\eta(\beta)=max(0,\beta)$.

\subsection{Random Forests (RF)}
Random Forests are a meta-modeling approach. They consist of ensembles or sets of decision trees \citet{panda2022machine,chung2021data}. Decision Trees (also known as Classification And Regression Trees) are simpler ML algorithms. Each decision tree partitions the feature or input space using inequalities. The various subsets of the partition are allotted specific values of the target. The decision trees are simple for use, however, due to the greedy nature of their training algorithm, these are prone to over-fitting. This can be eliminated with an ensemble of decor-related trees. The decorrelation can be prepared by bootstrapping the dataset and bagging the features. The estimations of the different trees can be aggregated to find the final prediction of the random forest (\citet{chung2020random}). The random forests have applications in different fields of engineering.    

Random forests have applications in different fields of engineering and science due to their accuracy, ease of use, and robustness. Random Forests have been applied to problems in renewable energy generation (\cite{serras2019combining}), estimating turbulence model parameters (\cite{heyse2021estimating}), prediction of meteorological coefficients (\cite{loken2019postprocessing}), simulation of combustion and chemical kinetics phenomena (\cite{chung2021interpretable}), etc. As a flexible model, Random Forests have many hyperparameters that can be changed depending on the complexity of the task at hand. The values of these hyperparameters need to be established to ensure a model that is generalizable. Generalizable models have the ability to have consistent performance in data that may not lie in their training set. The most important hyperparameters of RF are a number of decision trees in the ensemble $(n)$ and the maximum depth limiting each tree$(d)$. We focus our hyperparameter tuning on these two parameters.   

\subsection{Polynomial regression (PR)}
Polynomial regression (PR) is another approach to developing feature-target correlation. The advantage of polynomial regression over ANN is that the former can return interpretable equations that can easily be interpreted by human analysis. The targets are represented as a linear combination of input features (\citet{ostertagova2012modelling}). In the modeling of the correlations, we have used basis functions till second order. The second-order PR model can be defined mathematically as follows:    

\begin{equation}
y=\beta_0+ \sum_{i=1}^{k}\beta_ix_i+\sum_{i=1}^{k}\beta_{ii} x_i^2+ \sum_{i}\sum_{j}\beta_{ij}x_ix_{j}+\epsilon
\end{equation}

The $\beta_0$ term represents the intercept and captures constant baseline effects that are independent of the input features. The $\beta_i$  and $\beta_{ii}$ terms capture the linear and quadratic relationships between the input features and the target value. The $\beta_{ij}$ term represents the interactions amongst the input features. Thus the coefficients of this polynomial closure have clear human interpretability. Training of this model invokes gradient descent to find the appropriate coefficients for the dataset.
\begin{figure}
\begin{centering}
\includegraphics[width=0.4\textwidth]{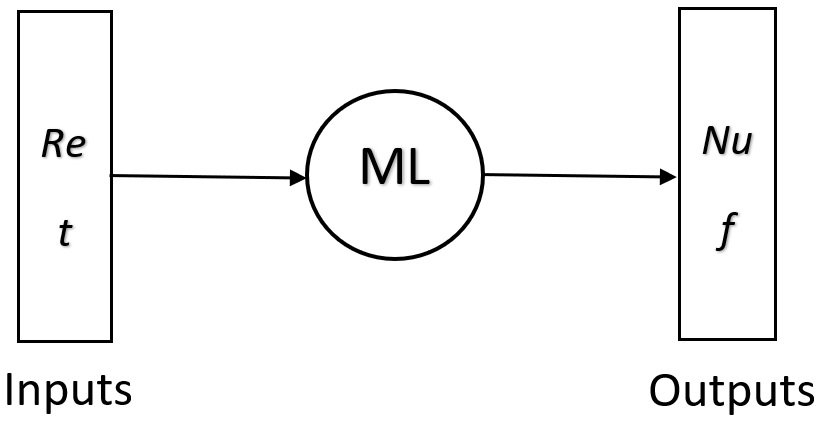}
\caption{ML block diagram representing features and levels.\label{fig:MLblock}}
\end{centering}
\end{figure}
\section{Training and testing of ML models}
The very first step of any ML model training and development is the fine-tuning of its hyperparameters. The hyperparameters are a set of parameters that affect the model development process and those must be optimized for finding the best prediction results from the ML models. The important hyper-parameters of random forest are a number of decision trees and a maximum depth of trees and for ANN, a number of layers and number of neurons in each layer. For polynomial regression, the degree of polynomial features can be changed to increase its effectiveness in predicting the quantities of interest. We have varied the hyperparameters of RF and the ANN as shown in table \ref{t_hyp}. For RF and ANN (a,b) correspond to (number of trees, maximum depth of trees) and (number of neurons in the first layer, number of neurons in the second layer) respectively. The variation of $R^2$ for different combinations of the hyperparameters is presented in figure \ref{fig:hypML}. In figure \ref{fig:hypML} a, b the evolution of $R^2$ in the testing of $Nu$ and $f$ are presented. From both the figures it is clear that $R^2$ values of ANN are greater than RF. The ANNs are universal approximators, that can correlate any type of level and features with minimal prediction error. The highest values of $R^2$ was noticed for RF and ANN hyper-parameters (20,25) and (30,30) respectively. However, the $R^2$ of ANN is much larger than the RF, for the testing results of $f$.
The $R^2$ of ANN predictions is much closer to unity for both the cases of $Nu$ and $f$ predictions.
\section{Predictions of unknown $Nu$ and $f$}
\begin{figure}
\begin{centering}
\subfloat[]{\includegraphics[width=0.4\textwidth]{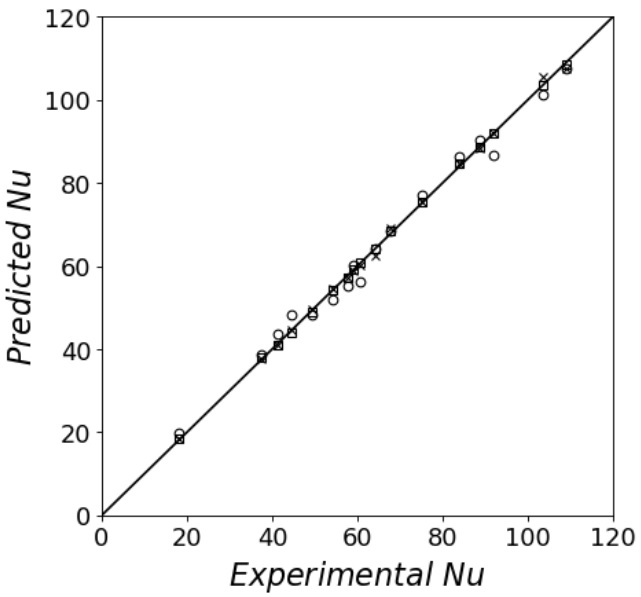}}
\subfloat[]{\includegraphics[width=0.4\textwidth]{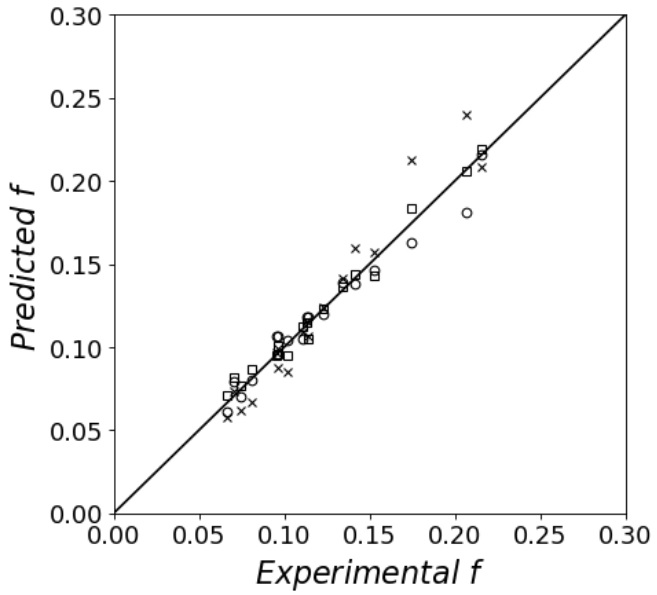}}
\caption{Experimental vs predicted values Nu and f. Circles, squares and cross symbols correspond to predictions of ANN, PR and RF respectively for train-test dataset $D_1$.\label{fig:NU_f_R}}
\end{centering}
\end{figure}

\begin{figure}
\begin{centering}
\subfloat[]{\includegraphics[width=0.4\textwidth]{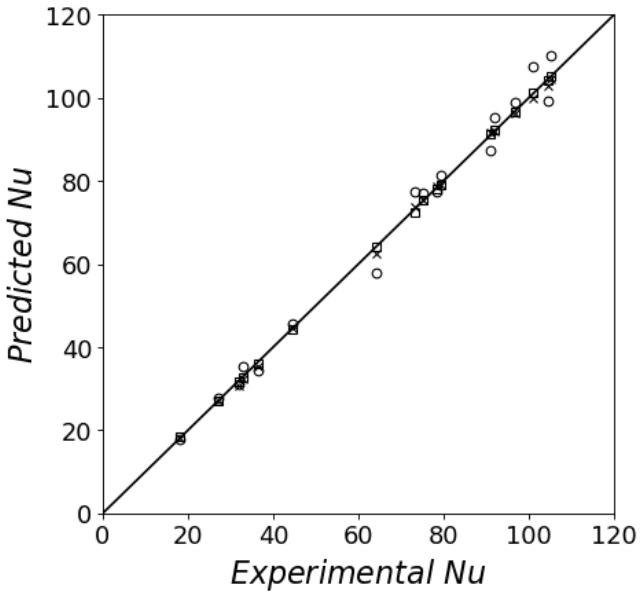}}
\subfloat[]{\includegraphics[width=0.4\textwidth]{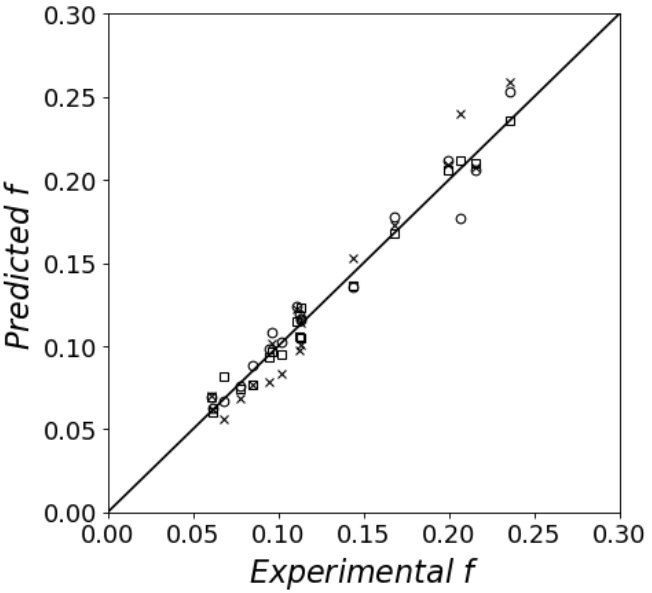}}
\caption{Experimental vs predicted values Nu and f. Circles, squares and cross symbols correspond to predictions of ANN, PR and RF respectively for train-test dataset $D_2$.\label{fig:NU_f_t}}
\end{centering}
\end{figure}

It is usual practice in ML based modeling is to keep some data for prediction purposes. We have kept few data for the predicting the $Nu$ and $f$ for unknown values of $Re$ and $t$ (these data were not used in the model development). In figure \ref{fig:NU_f_R} and \ref{fig:NU_f_t} the prediction results of ML based surrogates are presented. In figure \ref{fig:NU_f_R} and \ref{fig:NU_f_t} the $Nu$ and $f$ are plotted against $Re$ and $t$ respectively. For figure \ref{fig:NU_f_R} the $R^2$ and $MSE$ of predictions are presented in table \ref{t_R2_pred}. In table \ref{t_R2_pred} case 1 and 2 correspond to prediction results for dataset $D_1$ and $D_2$ respectively. For both the cases, in the prediction of $Nu$, the ANN models has the highest $R^2$ value. For case 1 and 2 the $R^2$ values was found to be 0.9999 and 0.9987 respectively. Similarly in the prediction of $f$ for both case 1 and 2, the maximum $R^2$ correspond to ANN. The $R^2$ values are 0.9685 and 0.9614 respectively. 
Although the ANN models have predicted the $Nu$ and $f$ accurately, the correlations developed through ANN are not interpretable. Hence we have presented the correlation learned through polynomial regression in equation form:  

\begin{figure}
\begin{centering}
\subfloat[$Nu$]{\includegraphics[width=0.4\textwidth]{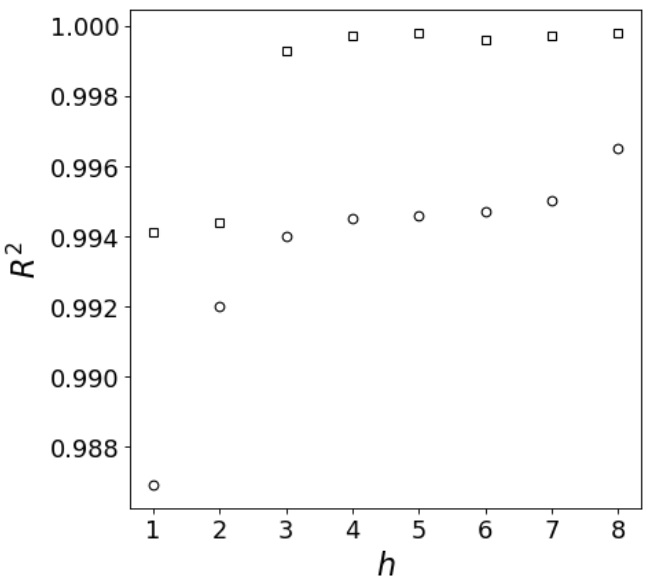}}
\subfloat[$f$]{\includegraphics[width=0.4\textwidth]{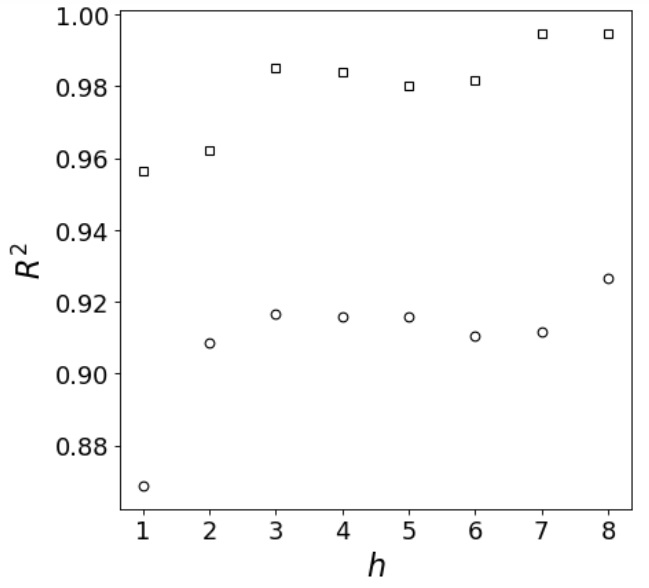}}
\caption{Hyperparameter optimization of RF and ANN. For RF and ANN 1, 2, 3, 4, 5, 6, 7, 8 correspond to hyperparameter sets as described in table \ref{t_hyp}. The square and circle correspond to ANN and RF respectively. \label{fig:hypML}}
\end{centering}
\end{figure}

\begin{table*}
\begin{center}
\begin{tabular} {c c c c c c c c c} 
  \toprule
     \hline
  \bfseries Models &  \multicolumn{1}{c}{\bfseries $1$} & \multicolumn{1}{c}{\bfseries $2$} & \multicolumn{1}{c}{\bfseries $3$} & \multicolumn{1}{c}{\bfseries $4$} &
  \multicolumn{1}{c}{\bfseries $5$} & \multicolumn{1}{c}{\bfseries $6$} & \multicolumn{1}{c}{\bfseries $7$} & \multicolumn{1}{c}{\bfseries $8$}\\
  \midrule
  RF (a,b) &  2,5 &  6,5 & 4,10  & 6,10 &  6,15 &  8,15 & 10,15  & 20,25 \\
ANN (a,b) & 2,2  & 5,5  & 8,8  & 12,12 &  16,16 &  20,20 & 26,26  & 30,30 \\
\bottomrule
\end{tabular}
\caption{The hyperparameter sets used in the optimization process. for RF and NN (a,b) correspond to (number of trees, maximum depth) and (number of neurons in first layer, number of neurons in second layer) respectively.}\label{t_hyp}
\end{center}
\end{table*}




\begin{figure}
\begin{centering}
\subfloat[]{\includegraphics[width=0.4\textwidth]{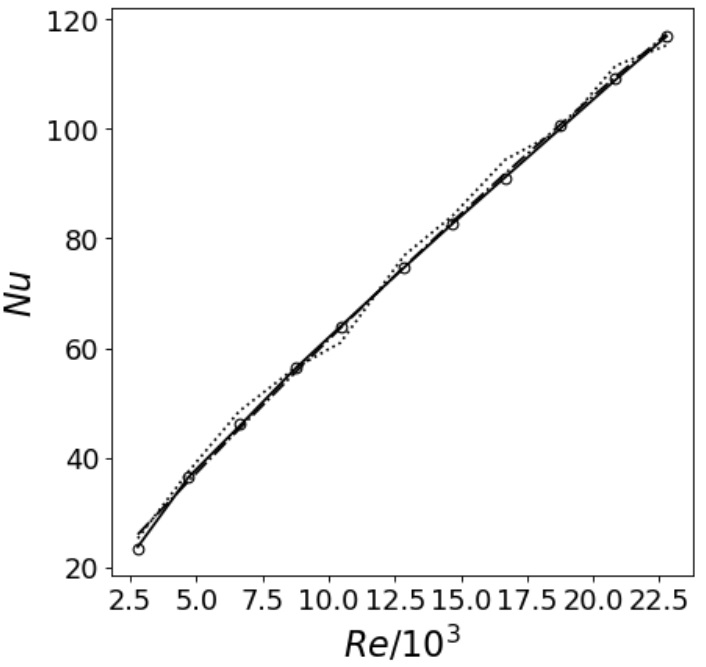}}
\subfloat[]{\includegraphics[width=0.41\textwidth]{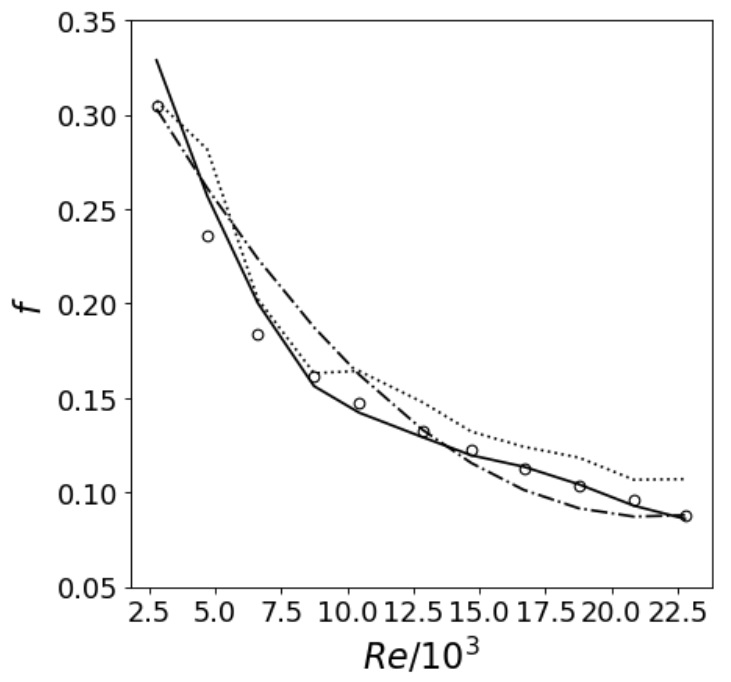}}
\caption{ML assisted prediction of $Nu$ and $f$ for different $Re$. Solid, dashed-dot and dotted lines correspond to predictions of ANN, PR and RF respectively for dataset $D_1$.\label{fig:NU_f_R_prediction}}
\end{centering}
\end{figure}
\begin{figure}
\begin{centering}
\subfloat[]{\includegraphics[width=0.4\textwidth]{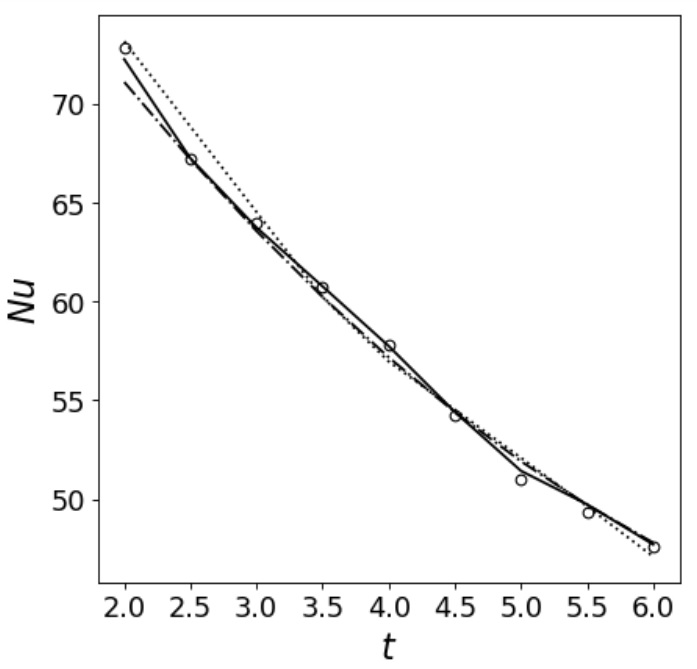}}
\subfloat[]{\includegraphics[width=0.425\textwidth]{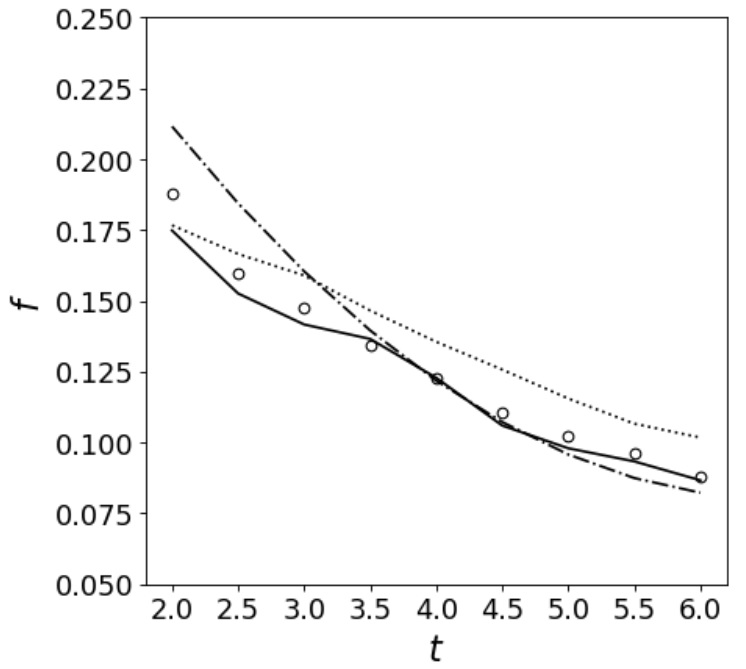}}
\caption{ML assisted prediction of $Nu$ and $f$ for different $t$. Solid, dashed-dot and dotted lines correspond to predictions of ANN, PR and RF respectively for for dataset $D_2$\label{fig:NU_f_t_prediction}}
\end{centering}
\end{figure}
\begin{table}
\begin{center}
\begin{tabular} {c c c c c } 
  \toprule
  \textbf{Case 1}
  \bfseries  &  \multicolumn{2}{c}{\bfseries $Nu$}  &  \multicolumn{2}{c}{\bfseries $f$}\\
     \hline
  \bfseries Models &  \multicolumn{1}{c}{\bfseries $R^2$} & \multicolumn{1}{c}{\bfseries $MSE$} & \multicolumn{1}{c}{\bfseries $R^2$} & \multicolumn{1}{c}{\bfseries $MSE$}\\
  \midrule
  ANN &  0.9998 &  0.1035 & 0.9840  & 2.87e-05  \\
PR & 0.9989  & 0.7307  & 0.8831  & 0.0002 \\
  RF & 0.9904 & 6.6531 & 0.9592  & 7.3627\\

\toprule
  \textbf{Case 2}
  \bfseries  &  \multicolumn{2}{c}{\bfseries $Nu$}  &  \multicolumn{2}{c}{\bfseries $f$}\\
     \hline
  \bfseries Models &  \multicolumn{1}{c}{\bfseries $R^2$} & \multicolumn{1}{c}{\bfseries $MSE$} & \multicolumn{1}{c}{\bfseries $R^2$} & \multicolumn{1}{c}{\bfseries $MSE$}\\
  \midrule
  ANN & 0.9999 &  0.0927 & 0.9856  & 4.2267  \\
PR & 0.9992  & 0.6757  & 0.9372  & 0.0001 \\
  RF & 0.9870 & 12.288 & 0.9602  & 0.0001\\
\bottomrule
\end{tabular}
\caption{$R^2$ and $MSE$ of testing by optimized ML models. Case 1 and 2 correspond to dataset $D_1$ and $D_2$ respectively}\label{t_R2_test}
\end{center}
\end{table}

\begin{table}
\begin{center}
\begin{tabular} {c c c c c } 
  \toprule
  \textbf{Case 1}
  \bfseries  &  \multicolumn{2}{c}{\bfseries $Nu$}  &  \multicolumn{2}{c}{\bfseries $f$}\\
     \hline
  \bfseries Models &  \multicolumn{1}{c}{\bfseries $R^2$} & \multicolumn{1}{c}{\bfseries $MSE$} & \multicolumn{1}{c}{\bfseries $R^2$} & \multicolumn{1}{c}{\bfseries $MSE$}\\
  \midrule
  ANN & 0.9999 & 0.1684 & 0.9685  & 2.09e-05  \\
PR & 0.9988  & 0.9837  & 0.9201  & 0.00031 \\
  RF & 0.9951 & 4.2084 & 0.9115  & 0.00035\\

\toprule
  \textbf{Case 2}
  \bfseries  &  \multicolumn{2}{c}{\bfseries $Nu$}  &  \multicolumn{2}{c}{\bfseries $f$}\\
     \hline
  \bfseries Models &  \multicolumn{1}{c}{\bfseries $R^2$} & \multicolumn{1}{c}{\bfseries $MSE$} & \multicolumn{1}{c}{\bfseries $R^2$} & \multicolumn{1}{c}{\bfseries $MSE$}\\
  \midrule
  ANN & 0.9987 & 0.0848 & 0.9614  & 0.00005  \\
PR & 0.9919  & 0.5315  & 0.8218  & 0.00017 \\
  RF & 0.9901 & 0.6192 & 0.8477  & 0.00014\\
\bottomrule
\end{tabular}
\caption{$R^2$ and $MSE$ of predictions by optimized ML models. Case 1 and 2 correspond to dataset $D_1$ and $D_2$ respectively}\label{t_R2_pred}
\end{center}
\end{table}
\begin{equation}
\begin{split}
&Nu=67.36-42.95Re-12.92t-3.04Re^2-5.45Re*t+2.27t^2,\\
&f=0.1-0.09Re-0.06t+0.06Re^2+0.04Re*t+0.03t^2,
\end{split}
\end{equation}

\section{Conclusions}
In this article, ML-based correlations or surrogate models were developed for Nusselt number and friction factor for a cylindrical heat exchanger with twisted tape inserts. The experimental heat transfer data at different Reynolds numbers were used for the ML model development. The dataset was divided into two different train-test datasets by taking the testing data with different combinations of Reynolds number and twist ratio. The first test data has all variations of Reynolds numbers and the fixed twist ratio of 3 and the second test data has all twist ratio variations and the fixed Reynolds number of 10449. From testing and prediction Results of the Nusselt number and friction factor, it is clear that the ANN predictions are better in comparison to the PR and RF. In most cases of Nusselt number of predictions is greater than or equal to 0.9987. However the $R^2$ value lies in the range of 0.9614 to 0.9856 , for the prediction of friction factor. The lowest value of $R^2$ corresponds to dataset $D_2$ for friction factor prediction at different twist ratios for fixed Reynolds number. Since in most cases of heat transfer correlation predictions using ANN, the $R^2$ magnitude is close to 1, we recommend the use of ANN for modeling the heat transfer correlations for heat exchangers with twisted tape inserts. In the future course of work, the correlation developed through polynomial regression can be used for the development of a complex objective function, that can be used for shape optimization of the twisted tape using meta-heuristic algorithms.     

\bibliography{asme2e.bib}
\end{document}